\documentclass[twocolumn,showpacs,preprintnumbers,amsmath,amssymb,prb]{revtex4}

\usepackage{graphicx}
\usepackage{dcolumn}
\usepackage{natbib}

\begin{document}
\bibliographystyle{plainnat}

\preprint{}

\title{
Dynamics and Spectral Weights of Shake-up Valence Excitations in
Transition Metal $K$-edge Resonant Inelastic X-ray Scattering}

\author{ K. H. Ahn,$^1$ A. J. Fedro,$^{2,3}$ and Michel van Veenendaal$^{2,3}$}
\affiliation{$^1$Department of Physics, New Jersey Institute of Technology, Newark, New Jersey, 07102 \\
$^2$Advanced Photon Source,
  Argonne National Laboratory, 9700 South Cass Avenue, Argonne,
Illinois 60439  \\
$^3$Department of Physics, Northern Illinois University,
De Kalb, Illinois 60115 }%


\begin{abstract}
Using a model Hamiltonian, we discuss how we could interpret data
obtained from transition metal $K$-edge resonant inelastic x-ray
scattering (RIXS) experiments.
By analyzing the creation of valence excitations
from the screening of the core hole and calculating corresponding RIXS spectra
for metals and insulators,
we find that the probability for excitations depends not only on the total energy but also
on the asymmetric screening dynamics between electrons and holes.
\end{abstract}


\maketitle

\section{Introduction}
\begin{figure}
\begin{center}
\includegraphics[width=8.6cm]{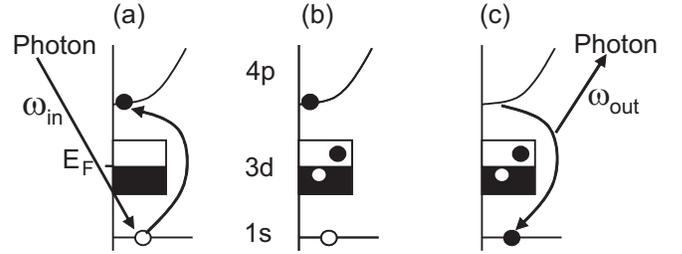}
\end{center}
\caption{\label{fig:krixs} Illustration of the RIXS process studied in our
work, that is, transition metal oxide $K$-edge RIXS.
This process is distinguished
from other types of RIXS, such as transition metal oxide
$L$-edge RIXS (Figures~1 and 2 in Ref.~\onlinecite{Gelmukhanov98}), or
$K$-edge RIXS for diamond or graphite(Figure~3 in
Ref.~\onlinecite{MvVgraphite}). }
\end{figure}

The inelastic x-ray scattering cross section from a solid state
system is enhanced by orders of magnitude, when the incoming
x-ray energy $\omega_{\rm in}$ is tuned to a resonance of the
system.
For resonant inelastic x-ray scattering (RIXS),
an electron is excited from a deep-lying core state into the valence shell.
In particular, the RIXS at transition metal $K$-edges,
which  is the focus of this work,
 has drawn significant attention
 recently.~\cite{Kao,Hill, Abbamonte,Hasan,Kim,Doring,Forte,Saitoh,Kondo,Grenier}
As illustrated in Fig.~\ref{fig:krixs},
transition metal $K$-edge RIXS  involves
 a dipolar transition from the $1s$ shell into the wide $4p$ band [Fig.~\ref{fig:krixs}(a)],
the screening of the $1s$ core hole by the $3d$ valence
electrons [Fig.~\ref{fig:krixs}(b)], and
the recombination of the $4p$ electron and the $1s$ core hole
with the emission of a photon [Fig.~\ref{fig:krixs}(c)].
Unlike RIXS at other edges
(e.g. transition-metal $L$-edges), the dipole matrix elements in
transition metal $K$-edge RIXS at a particular incoming photon
energy ($\omega_{\rm in}$) and momentum (${\bf q}_{\rm in}$) are
constant factors, which makes transition metal $K$-edge RIXS a
direct probe for the screening dynamics.
The momentum (${\bf q}_{\rm out}$) and energy ($ \omega_{\rm out}$)
dependence of the outgoing x-ray resulting from the radiative decay of the core hole allows
measurements of the energy $\omega=\omega_{\rm in}-\omega_{\rm out}$ and
the momentum ${\bf q}={\bf q}_{\rm in}-{\bf q}_{\rm out}$ of
elementary excitations near the Fermi level or across the gap.
The RIXS has
several advantages compared to other spectroscopic probes. In
contrast to angle-resolved photoemission, RIXS is bulk-sensitive
due to the large penetration depth of x-rays. The use of a
particular resonance makes RIXS chemically selective.
As opposed to x-ray absorption, there is no core hole present in the
final state.
The RIXS has provided unique insights into, e.g., the
momentum dispersion of charge excitations and magnetic excitations in high-$T_c$
superconductors and related systems,~\cite{Hill,
Abbamonte,Hasan,Kim,Doring,Forte} orbital excitations,~\cite{Saitoh,Kondo}
and magnetic-ordering dependent transfer of
spectral weight in colossal magnetoresistive
manganites.~\cite{Grenier}

Numerical calculations of the RIXS
spectrum have been done on, e.g., finite
clusters.~\cite{Tsutsui,Nomura,Vernay}
In certain limits, the RIXS cross section can be related to
the dynamic structure factor $S_{\bf q}(\omega)$,
either by lowest order perturbation theory~\cite{Abbamonte,Doring} or by an
approximate representation of the intermediate-state
dynamics.~\cite{Brink}
Within these approximations, the probability for
the intermediate state core-hole potential to create excitations,
such as electron-hole (e-h) pairs, mainly depends on the total energy of
the shake-up process.
In this work, we show that RIXS depends on the energies of the constituents of the excitation,
leading to a strong asymmetry in electron and hole responses to the core hole
for the parameter range where the screening dynamics become important.
We  demonstrate that the RIXS spectral line shape is
determined by the nature of the intermediate-state core-hole
valence-shell-electron excitons.

\section{Model Hamiltonian and methods}
The total Hamiltonian for the RIXS process shown in Fig.~\ref{fig:krixs}
has electronic, photonic, and electron-photon interaction parts,
\begin{equation}
H_{\rm total}=H_{\rm el}+H_{\rm pht}+H_{\rm el-pht}.
\end{equation}
The RIXS intensity is obtained from
\begin{equation}
I(\omega_{\rm in},\omega_{\rm out})=\frac{2\pi}{\hbar} \sum_f \left| U_{i \rightarrow f} \right|^2
\delta \left( E_f + \omega_{\rm out} - E_i -\omega_{\rm in} \right),
\end{equation}
where $U_{i\rightarrow f}$ is
the transition amplitude from the initial state $|i\rangle$ to a final state $|f\rangle$.
In terms of
the initial many-electron energy $E_i$,
many-electron eigenstate $|n\rangle$ and eigenenergy $E_n$
of $H_{\rm el}$,
and the FWHM intermediate-state lifetime-broadening $2\Gamma$,
the transition amplitude is given by
\begin{equation}
U_{i \rightarrow f} = \sum_n
\frac{\left\langle f \left| H_{\rm el-pht} \right| n \right\rangle  \left\langle n \left| H_{\rm el-pht} \right| i \right\rangle }
{E_i+\omega_{\rm in}-E_n + i\Gamma}.
\label{eq:Uif}
\end{equation}

The Coulomb interaction between localized $1s$ holes and $3d$ electrons,
which is the central mechanism for transition metal $K$-edge RIXS,
is $U_c=$ 4 - 7 eV (Refs.~\onlinecite{Kim}, \onlinecite{Ide}, \onlinecite{Tsutsui}, and \onlinecite{Laan81}).
The excited $4p$ electron is highly delocalized,
and its Coulomb interaction with $1s$ holes or $3d$ electrons
is neglected as an approximation, leading to
\begin{equation}
H_{\rm el}=H_{\underline {1s}}+H_{3d}+H_{{\underline {1s}}-3d}+H_{4p},
\end{equation}
where $H_{\underline {1s}}$, $H_{3d}$, and $H_{4p}$ are Hamiltonians
for $1s$, $3d$ and $4p$ levels, and
$H_{{\underline {1s}}-3d}$ represents $1s$-$3d$ Coulomb interaction.
The $4p$ level Hamiltonian $H_{4p}$ is decoupled from the other terms,
and can be replaced with a constant $\varepsilon_{4p}$ for the intermediate state, which
we drop from $H_{\rm el}$.
At a particular resonant $\omega_{\rm in}$ for the $K$-edge RIXS,
the matrix element of $H_{\rm el-pht} \sim {\bf p}\cdot {\bf A}$
between $1s$ and $4p$ is a constant factor, leading to
\begin{eqnarray}
I(\omega_{\rm in},\omega_{\rm out})
&\propto&
\sum_f \left| \sum_{n}
\frac{ \left\langle f \left| {\underline s} \right| n \right\rangle  \left\langle n \left| {\underline s}^{\dagger} \right| i \right\rangle}
{E_i+\omega_{\rm in}-E_n + i\Gamma}
 \right|^2   \nonumber \\
&&\delta \left( E_f + \omega_{\rm out} - E_i -\omega_{\rm in} \right), \label{eq:Iomega}
\end{eqnarray}
where ${\underline s}^{\dagger}$ is a $1s$ hole creation operator.
Since transition metal $K$-edge RIXS probes
initial and final states without core holes and
intermediate states with core holes,
$|i\rangle$ and $|f\rangle$ are the energy eigenstates of
$H_{\rm el}$ with  ${\underline s}^{\dagger} {\underline s} = 0$,
that is, $H_{3d}$,
whereas $|n\rangle$ are the energy eigenstates of
$H_{\rm el}$ with  ${\underline s}^{\dagger} {\underline s} = 1$.
Equation~(\ref{eq:Iomega}) is evaluated by
expanding $3d$ parts of $|n\rangle$
in the basis of the energy eigenstates of $H_{3d}$,
which include $3d$ parts of $|i\rangle$ and $|f\rangle$.

In terms of a $3d$ electron creation operator $d^{\dagger}$,
the $3d$ Hamiltonian for metals and insulators is
\begin{equation}
H_{3d}^{\rm m/i}=\sum_{k} {\varepsilon}_{k}
{d}^{\dagger}_{k} {d}_{k},
\end{equation}
for which  we consider a constant DOS without and with a gap ($\Delta$) for metals and insulators, respectively.
The Hamiltonian includes
$N$ one-electron $3d$ levels.
The number of $3d$ electrons is $L$.
For independent electrons with a single orbital degree of freedom at each site,
the two spin channels are equivalent, and two bound
states are present at the site with the core hole.
However, for many transition-metal compounds, the Coulomb
interaction between valence electrons is often non-negligible,
resulting in the Mott insulators.
For these Mott insulators,
we consider the Hubbard Hamiltonian,
\begin{equation}
H_{3d}^{\rm Hubbard}=\sum_{k,\sigma=\uparrow,\downarrow} {\varepsilon}_{k}
{d}^{\dagger}_{k,\sigma} {d}_{k,\sigma}
+U\sum_i d^\dagger_{i\uparrow}
d_{i\uparrow}d^\dagger_{i\downarrow} d_{i\downarrow},
\end{equation}
which includes the effect of the on-site $3d$-$3d$ Coulomb interaction.
Spin and site indices are $\sigma$ and $i$, respectively.

Because the $1s$ band is much narrower than the $3d$ band,
the total RIXS intensity is, in a good approximation, proportional to
the intensity calculated for a $1s$ hole created at, for example, site $i=0$ (see footnote~\onlinecite{local.approx}).
Therefore, we use $1s$ hole Hamiltonian,
\begin{equation}
H_{\underline {1s}}=\varepsilon_{1s} {\underline s}^{\dagger} {\underline s}.
\end{equation}
The following identity shows that the on-site Coulomb interaction with a localized core hole scatters
any $3d$ state with a wavevector $k$ into the same or any other $3d$ state with a wavevector $k'$
with an equal probability,
\begin{equation}
-U_c d^{\dagger}_{i=0} d_{i=0} {\underline s}^{\dagger}_{i=0} {\underline s}_{i=0}
= - \sum_{k,k'} \frac{U_c}{N} d^{\dagger}_{k} d_{k'} {\underline s}^{\dagger}_{i=0} {\underline s}_{i=0}.
\end{equation}
 For our model, this corresponds to
the scattering from any $3d$ level into the same or any other $3d$ level
with the same strength of $-U_c/N$,
which is expressed in the Hamiltonian,
\begin{equation}
H_{\underline {1s}-3d}= - \sum_{k,k'}
\frac{U_c}{N} d^{\dagger}_{k} {d}_{k'} {\underline s}^{\dagger} {\underline s}.
\end{equation}

Our model Hamiltonian for metals and insulators with the core hole, that is,
$H_{\rm el}=H_{3d}^{\rm m/i}+H_{\underline{1s}-3d}+H_{\underline {1s}}$
with ${\underline s}^{\dagger} {\underline s}=1$,
is an exactly-solvable independent-particle Hamiltonian, and has been solved in the context
of x-ray edge singularity.~\cite{Davis79, Davis80}
The intermediate $L$-electron eigenstates and eigenenergies are obtained
by diagonalizing $H_{\rm el}$ exactly and
filling $L$ one-electron eigenstates.
To calculate the RIXS intensity, we classify final states
according to the number of e-h pairs, for example,
the Fermi sea (i.e., the initial state $|i\rangle$),
states with one e-h pair $d_{k_>}^{\dagger}{d}_{k_<} |i\rangle$, and
states with two e-h pairs $d_{k_>}^{\dagger}d_{k'_>}^{\dagger}{d}_{k_<} {d}_{k'_<} |i\rangle$,
where $k_<$ and $k_>$ represent
$3d$ electron levels below and above the Fermi energy, respectively.
We examine the contribution of each class of states
to the RIXS intensity, as we increase the number of e-h pairs
in the final states.

For the Mott insulators,
the Hubbard model cannot be solved exactly.
However, it is known that the main features of the Mott insulators are
understood by following Hubbard's own approximation~\cite{Hubbard} for $H_{3d}^{\rm Hubbard}$,
which we extend to
calculate the RIXS spectral line shape for the half-filled Mott insulators.
First, we project the
operator for a particular spin onto different occupations of the opposite
spin,
\begin{equation}
d^\dagger_{i\sigma}=(1-n_{i,-\sigma}) d^\dagger_{i\sigma}+
n_{i,-\sigma} d^\dagger_{i\sigma}.
\end{equation}
The first operator (singlon)
$g^\dagger_{i\sigma}=(1-n_{i,-\sigma}) d^\dagger_{i\sigma}$
describes the motion of the electron in
the absence of the electron with the opposite spin. The second operator (doublon)
$b^\dagger_{i\sigma}=n_{i,-\sigma} d^\dagger_{i\sigma}$ describes the motion
of the electron when the site is doubly occupied. The local potential in the presence of the
core hole at $i=0$ is written as
\begin{equation}
H_{\underline{1s}-3d}= - \sum_{\sigma} U_c \left(g^\dagger_{i=0,\sigma}g_{i=0,\sigma}+b^\dagger_{i=0,\sigma}b_{i=0,\sigma} \right ).
\label{coreHubbard}
\end{equation}
In the Hubbard-I approximation,~\cite{Hubbard}
the projection operators, $1-n_{i,-\sigma}$ and  $n_{i,-\sigma}$, are replaced by their expectation values,
resulting in an effective independent-particle Hamiltonian.
The one-particle eigenvectors for the lower and
upper Hubbard bands in the absence of the core-hole potential
are written as
\begin{eqnarray}
|\varepsilon_{{\bf k}L\sigma}\rangle &=&\cos \theta_{\bf k} |g_{{\bf k}\sigma}\rangle
+\sin \theta_{\bf k} |b_{{\bf k}\sigma}\rangle, \\
|\varepsilon_{{\bf k}U\sigma}\rangle &=&\sin \theta_{\bf k} |g_{{\bf k}\sigma}\rangle
-\cos \theta_{\bf k} |b_{{\bf k}\sigma}\rangle,
\end{eqnarray}
respectively, where {\bf k} is the wavevector.
Although the core-hole potential in Eq.~(\ref{coreHubbard})
does not couple singlon and doublon states, scattering between the
lower and upper Hubbard bands still occurs due to the mixing
of the singlon and doublon states through electron hopping.
In the initial state, only the states in the lower Hubbard band are occupied,
\begin{eqnarray}
|i\rangle=\prod_{{\bf k}\sigma} |\varepsilon_{{\bf k}L\sigma}\rangle.
\end{eqnarray}
For the intermediate states,
the effective independent-particle Hamiltonian is solved
in the presence of the core-hole potential, leading to
one-particle eigenstates $|\varepsilon_{m\sigma}\rangle$, where the index $m=1,2,3,\cdots$ labels
the eigenstates in the order of increasing energy.
We note that the wavevector ${\bf k}$ is no longer a good quantum
number and that the lower and upper Hubbard bands are no longer
well defined because of the presence of a bound state at the
site with the core hole.
Many-electron intermediate eigenstates and energies are obtained by
filling an appropriate number of the states, $|\varepsilon_{m\sigma}\rangle$.
By following the same procedure as for metals and insulators,
we find the RIXS spectral shape for the Mott insulators.

\section{Results}
\begin{figure}
\begin{center}
\includegraphics[width=5.0cm]{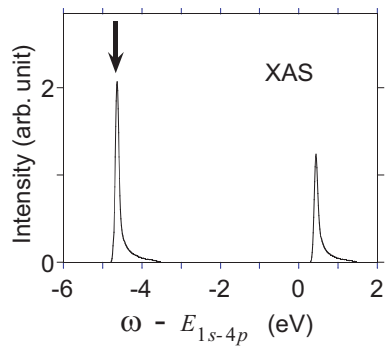}
\end{center}
\caption{\label{fig:xas}
X-ray absorption spectroscopy (XAS) spectrum for
a metallic DOS with a bandwidth $W$ = 2 eV and a core-hole potential $U_c$ =  5 eV
at half-filling.}
\end{figure}

To find RIXS spectrum,
all possible intermediate states should be considered in principle,
as done for small clusters in Refs.~\onlinecite{Tsutsui} and~\onlinecite{Nomura}.
However, for most RIXS experiments, the incoming x-ray energy is tuned at the absorption threshold
for the resonance enhancement effect.
Therefore, without the short lifetime of the intermediate state and the energy uncertainty,
the intermediate state would be purely the lowest energy state,
in which the localized core hole is screened by the $3d$ electrons.
To find the contribution of different intermediate states to the RIXS intensity,
we calculate a part of the RIXS matrix element
$\langle n|{\underline s}^{\dagger} |i\rangle$ in Eq.(\ref{eq:Iomega}),
namely the projection of the intermediate states onto the state of the Fermi sea with a core hole,
which corresponds to the x-ray absorption intensity,
\begin{equation}
I_{\rm XAS}(\omega) \propto \sum_{n} \left| \langle n|{\underline s}^{\dagger} |i\rangle  \right|^2 \delta(\omega-E_n+E_i).
\end{equation}
The x-ray absorption intensity, calculated
for a square density of states with a band width $W$ = 2 eV and $U_c$ = 5 eV at half-filling,
is displayed in Fig.~\ref{fig:xas}, which shows
two distinct spectral bands corresponding
to the XAS final states with an almost full (screened) and an almost empty (unscreened) bound
state, separated by approximately $U_c$ (Ref.~\onlinecite{Davis79}).
Experimentally, the energy difference between the onsets of resonant inelastic x-ray
scattering for a screened and an unscreened core hole is 5-7 eV (Ref.~\onlinecite{Kim}),
which corresponds to $U_c$.
For the lower spectral band, the XAS spectral weight is dominated by the
transitions into the lowest energy intermediate state, $|n_{\rm low}\rangle$,
whereas the higher-lying states in this band
have little component of
the Fermi sea with a core hole, ${\underline s}^{\dagger} |i \rangle $.
It shows that the lowest energy screening state, indicated by an arrow in Fig.~\ref{fig:xas}, dominates the matrix element.
The smaller satellite peak at higher energy corresponds to the excitation of
the unscreened state.
Since the core-hole lifetime broadening $\Gamma$ is significantly smaller than the
energy difference between the screened and unscreened absorption states,
the probability for the unscreened state excitation would be low.
Even if the unscreened state is partially excited,
its matrix element for RIXS is smaller than that for the screened state.
Based on these arguments, we propose that the screened state is
the dominant intermediate state contributing to the RIXS intensity, and we replace
the sum over the intermediate states with a single lowest
energy intermediate state, $n_{\rm low}$:
\begin{equation}
I(\Delta \omega) \propto \sum_f \left| \left\langle f | {\underline s} | n_{\rm low} \right\rangle \right|^2
\delta[\Delta \omega -(E_f - E_i)]. \label{eq:Igen}
\end{equation}

For metals and insulators,
using indices $k$ and $m$, which are reserved for
$3d$-level one-electron eigenstates without and with the core hole respectively throughout this paper,
we express operators for the intermediate one-particle eigenstates with energy $\varepsilon_m$
($m=1,2,...,N$ in the order of increasing energy) as
\begin{equation}
D_m^{\dagger}=\sum_k a_{mk} d_k^{\dagger}. \label{eq:amk}
\end{equation}
The numerically exact lowest energy intermediate state $|n_{\rm low}\rangle$
is found
by filling up the lowest $L$ one-electron energy eigenstates of
$H_{\rm el}$ with ${\underline s}^{\dagger} {\underline s}=1$.

\begin{figure}
\begin{center}
\includegraphics[width=8.6cm]{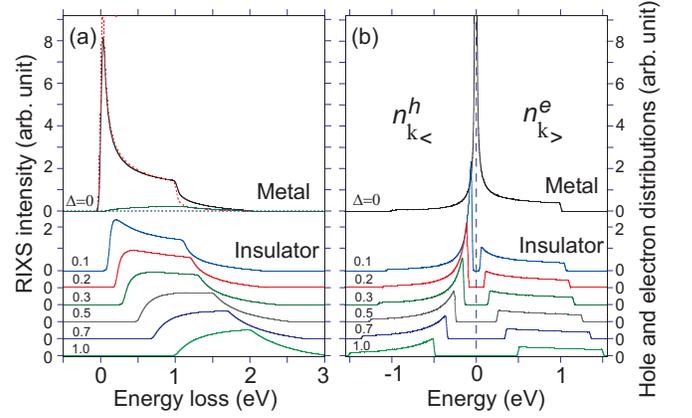}
\end{center}
\caption{\label{fig:mi_rixs} (color online) (a) RIXS spectra calculated for a square density of
states with a band width $W=2$ eV and $U_c=5$ eV at half-filling for a metal [upper (black) solid line for $\Delta=0$]
and insulators ($\Delta > 0$).
For the insulators, a gap ($\Delta$) is opened at the center of the band, and the occupied and unoccupied
bands have the same width $W/2$.
For the metal ($\Delta=0$), the lower (green) solid line shows
the contribution of the double electron-hole pairs to the RIXS
spectrum and the upper (red) dotted line gives the approximate
spectrum using Eq.~(\ref{eq:Inenh}).
The lower (green) dotted line is a shifted energy axis.
(b) Distribution functions of holes ($n_{{k}_<}^h$) and
electrons ($n_{{k}_>}^e$) below and above the Fermi level for different values of $\Delta$,
created by the intermediate-state core-hole
potential.}
\end{figure}

The RIXS intensity in Eq.~\ref{eq:Igen} calculated for the final states with up to three e-h pairs
for a half-filled metallic DOS is shown by
the upper (black) solid line marked with $\Delta=0$ in Fig.~\ref{fig:mi_rixs}(a).
The full bandwidth and the core-hole potential
are chosen as $W=2$ eV and $U_c=5$ eV, respectively.
The RIXS intensity calculated only for the final states with two e-h pairs is shown
by the lower (green) solid line marked with $\Delta=0$ in Fig.~\ref{fig:mi_rixs}(a),
which is small compared to the total intensity.
The contribution of the final states with three e-h pairs to the RIXS intensity is negligible.
The results show that the final states with one e-h pair contribute
dominantly to the RIXS intensity,
which means that the lowest intermediate state
is approximately a combination of the Fermi sea and the states with one e-h pair,
\begin{equation}
|n_{\rm low}\rangle \approx \alpha_{\rm low}{\underline s}^{\dagger}|i\rangle+\sum_{k_>,k_<}\beta_{\rm low}(k_>,k_<)
d^{\dagger}_{k_>} d_{k_<} {\underline s}^{\dagger}|i\rangle.
\end{equation}
The first term gives rise to a large $\langle n_{\rm low}|{\underline s}^{\dagger} |i\rangle$,
the XAS part in the RIXS matrix element in Eq.~(\ref{eq:Iomega}) as discussed earlier.
The remaining terms with one e-h pair give
a large $\langle f|{\underline s}|n_{\rm low}\rangle$, the other part in the RIXS matrix element.

The electron and the hole states in $|n_{\rm low}\rangle$ are entangled in general.
We examine whether unentangled electron and hole states serve as an approximation,
that is,
\begin{eqnarray}
&&|n_{\rm low}\rangle \approx \alpha_{\rm low}{\underline s}^{\dagger}|i\rangle+ \nonumber \\
&&\left( \sum_{k_>}\beta^e_{\rm low}(k_>) d^{\dagger}_{k_>} \right)
\left( \sum_{k_<}\beta^h_{\rm low}(k_<) d_{k_<} \right) {\underline s}^{\dagger} |i\rangle. \label{eq:one_pair}
\end{eqnarray}
We calculate the electron and hole distribution functions for the lowest energy intermediate
state using $a_{mk}$ defined in Eq.~(\ref{eq:amk}),
\begin{eqnarray}
n^{e}_{k_>}(\varepsilon_{k_>})&=&\sum_{m=1,L} | a_{mk_>} |^2, \label{eq:ne} \\
n^{h}_{k_<}(\varepsilon_{k_<})&=&\sum_{m=L+1,N} | a_{mk_<} |^2, \label{eq:nh}
\end{eqnarray}
which are shown in Fig.~\ref{fig:mi_rixs}(b) for $\Delta=0$.
If Eq.~(\ref{eq:one_pair}) is exact,
we would have
\begin{eqnarray}
n^{e}_{k_>}(\varepsilon_{k_>}) &\propto & | \beta^e_{\rm low}(k_>) |^2, \\
n^{h}_{k_<}(\varepsilon_{k_<}) &\propto & | \beta^h_{\rm low}(k_<) |^2
\end{eqnarray}
and
\begin{eqnarray}
I(\Delta \omega)=A \sum_{k_>,k_<}
n^e_{k_>}(\varepsilon_{k_>}) n^{h}_{k_<}(\varepsilon_{k_<})
\delta (\Delta \omega-\varepsilon_{k_>}+\varepsilon_{k_<}). \label{eq:Inenh}
\end{eqnarray}
Therefore, the comparison of
the RIXS intensity
calculated from Eq.~(\ref{eq:Inenh})
with the one calculated from Eq.~(\ref{eq:Igen}) reveals
whether Eq.~(\ref{eq:one_pair}) is a good approximation.
The RIXS spectrum calculated from Eq.~(\ref{eq:Inenh}) is shown
with the upper (red) dotted line for $\Delta=0$
in Fig.~\ref{fig:mi_rixs}(a), which
shows a good agreement with the upper (black) solid line,
indicating that a separable e-h pair state is a good approximation
for the lowest energy intermediate state.

We examine the characteristics of the electron and hole distribution functions,
$n^{e}(\varepsilon_{k_>})$ and $n^{h}(\varepsilon_{k_<})$, shown in Fig.~\ref{fig:mi_rixs}(b) for a metal ($\Delta$ = 0).
First, the distribution functions show a strong singularity near the Fermi level,
which is responsible for a similar singularity in the RIXS intensity, $I(\Delta \omega)$.
These features correspond to
the low energy e-h excitations right near the Fermi energy,
related to the x-ray edge singularity.~\cite{Mahan}
However, the elastic peak in RIXS overlaps with this edge singularity,
and higher energy features are more important.
Away from the Fermi energy,
hole spectral weight, $n^{h}_{k_<}(\varepsilon_{k_<})$, diminishes very rapidly,
whereas the electron spectral weight, $n^{e}_{k_>}(\varepsilon_{k_>})$,
remains nearly constant.
We interpret this difference in the following way.
In the lowest energy intermediate state, the electron screens the localized core hole,
meaning that the excited electron should be localized in the real space.
That corresponds to the state with a nearly constant amplitude in $k$-space,
or a constant amplitude in the energy space since we assume a constant electron DOS.
The excited $3d$ hole does not screen the core hole, but
is in a delocalized state right below the Fermi energy
to minimize its kinetic energy,
which corresponds to
the spectral weight concentrated near the Fermi energy.
Such asymmetry gives rise to the higher
energy features in RIXS.
The RIXS intensity for $\Delta = 0$ in Fig.~\ref{fig:mi_rixs}
is large up to 1 eV, i.e., the width of the unoccupied band,
beyond which the intensity quickly decreases to zero.
This asymmetry differs significantly from the
existing approaches in Refs.~\onlinecite{Abbamonte},~\onlinecite{Doring} and~\onlinecite{Brink}, which
predict that the RIXS spectral line shape is
proportional to the dynamic structure factor $S(\omega)$
(i.e., the convolution of the occupied and the unoccupied DOS) weighted by the
energy of the shake-up structure.
According to these studies,
the RIXS for a square DOS would be given by $S(\omega)$, which has a triangular
shape, multiplied by a $U_c^2/\omega^2$ (Refs.~\onlinecite{Abbamonte} and~\onlinecite{Doring}) or a
$U_c^2/(U_c-\omega)^2$ (Ref.~\onlinecite{Brink}) energy dependence.
However, Fig.~\ref{fig:mi_rixs}(a) clearly shows strong deviations from a simple triangular
line shape.
The results in Refs.~\onlinecite{Abbamonte} and~\onlinecite{Doring} are correct
in the limits of $U_c \rightarrow 0$ and $U_c\rightarrow \infty$, respectively.
For intermediate $U_c$, the screening dynamics becomes more important, leading to the
asymmetry in electron and hole excitations.

We also carry out similar calculations for insulators.
The results are shown in Fig.~\ref{fig:mi_rixs} for $\Delta > 0$.
For a small gap, the singularity disappears and the RIXS spectral line shape
resembles more closely
the unoccupied DOS than $S(\omega)$. For larger gap values of the
order of the bandwidth, the spectral line shape develops into a
triangular shape.
The electron-hole asymmetry between $n_{k_<}^h$ and $n_{k_>}^e$
becomes weaker, but persists.
It shows that the widths of RIXS peaks for metals
and insulators with gaps smaller than the band width
roughly correspond to the widths of unoccupied bands.
Therefore, RIXS probes mainly unoccupied bands,
complementing ARPES, which probes occupied bands.

\begin{figure}
\begin{center}
\includegraphics[width=8.6cm]{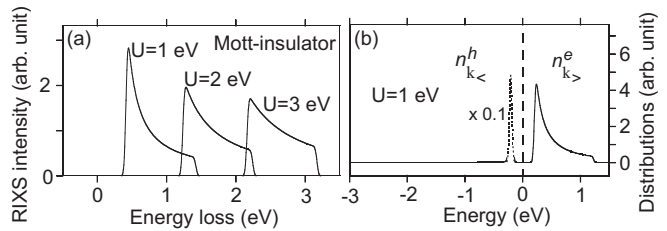}
\end{center}
\caption{\label{fig:mott_rixs} (color online)
(a) RIXS spectra calculated for the half-filled Mott insulators
with the on-site Coulomb interaction $U=1,2,$ and 3 eV ($U_c=5$ eV, $W=2$ eV).
(b) Distribution functions of holes and electrons created by the intermediate core-hole potential
for the Mott insulator with $U=1$ eV.
}
\end{figure}

The RIXS for the half-filled Mott insulators is calculated
by the same procedure used for metals and insulators.
For the lowest intermediate state, the lowest $N$ one-particle eigenstates are filled,
\begin{eqnarray}
|n_{\rm low}\rangle=\prod_{\sigma=\uparrow,\downarrow,m=1,\cdots,N/2} |\varepsilon_{m\sigma}\rangle.
\end{eqnarray}
De-excitations from the lowest intermediate state are obtained by
representing $|\varepsilon_{m\sigma}\rangle$ in terms of
$|\varepsilon_{{\bf k}L\sigma}\rangle$ and $|\varepsilon_{{\bf k}U\sigma}\rangle$,
and calculating the overlap with final e-h pair excitations, such as
\begin{eqnarray}
|f\rangle=|\varepsilon_{{\bf k_>}U\sigma'}\rangle
\prod_{
{\bf k}\sigma \ne {\bf k_<}\sigma'
}|\varepsilon_{{\bf k}L\sigma}\rangle.
\end{eqnarray}
The RIXS spectrum is calculated according to Eq.~(\ref{eq:Igen}).
The results for $U=1,2,$ and 3 eV
are shown in Fig.~\ref{fig:mott_rixs}(a),
where we consider final states with up to two e-h pairs.
Electron and hole distribution functions for the Mott insulators in Fig.~\ref{fig:mott_rixs}(b) show
even stronger asymmetry than those for metals and insulators.
The width of the RIXS peak is about $W/2$, which is the width of the unoccupied (upper Hubbard) band.
More discussions are given in the next section.

\section{Discussions}
We analyze the relations between the one-particle eigenstates with and without
the core hole to understand the results obtained in Sect.~III.
As mentioned in Sect.~II, the Hamiltonians with the core-hole potential have been studied in the context
of x-ray absorption edge singularity,~\cite{Davis79, Davis80} and it is known that
the coefficient relating the two basis sets in Eq.~(\ref{eq:amk}) is
\begin{equation}
a_{mk}=\frac{C_m}{\varepsilon_k-\varepsilon_m},
\end{equation}
where $C_m$ is a normalization factor and $\varepsilon_m$'s are the solutions of
\begin{equation}
\frac{1}{U_c}=\frac{1}{N}\sum_k \frac{1}{\varepsilon_k-\varepsilon_m}.
\end{equation}
Figure 1 in Ref.~\onlinecite{Davis79} illustrates how $\varepsilon_m$ is determined graphically
for a given set of $\varepsilon_k$'s and a value of $U_c$.

\begin{figure}
\begin{center}
\includegraphics[width=8.6cm]{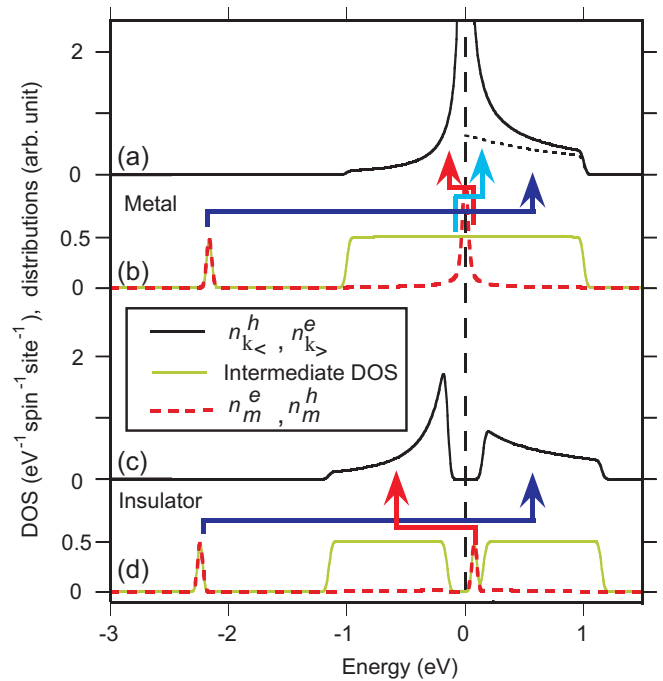}
\end{center}
\caption{\label{fig:mi_discuss}
(Color online)
(a) and (c) show the distribution functions of holes and electrons
in the intermediate state for a metal and an insulator ($\Delta=0.3$ eV) respectively,
similar to the ones shown in Fig.~\ref{fig:mi_rixs}(b).
(b) and (d) show the intermediate DOS with gray (green) solid lines
and $n^e_m$ and $n^h_m$, defined in the text, with (red) dashed lines.
Arrows indicate how the two plots are related, as explained in the text.
For illustrative purpose, a smaller core-hole potential ($U_c=2$ eV)
is chosen.
Other parameter values are identical to those for Fig.~\ref{fig:mi_rixs}. }
\end{figure}

First, we analyze the results for metals.
For a metallic DOS with $N$ states separated by an equal, small energy difference, the effect of
the strong attractive Coulomb potential $-U_c$ is to pull out a bound state below the bottom
of the band and shift the energies of the remaining $N-1$ states (see Fig. 1 in Ref.~\onlinecite{Davis79}).
The calculated DOS in the presence of the core-hole potential
is shown with the gray (green) solid line in Fig.~\ref{fig:mi_discuss}(b).
The bound state energy is $\varepsilon_{m=1}\cong -U_c-\frac{W^2}{12U_c}$.
The energies for the rest of the states are
$\varepsilon_m=\varepsilon_k-\frac{\delta_m}{\pi}\frac{W}{N}$,
where the indices $m$ and $k$ have the same value and  the phase shift
is $\delta_m=\pi,\dots,0$ for $\varepsilon_k=-\frac{W}{2},\dots,\frac{W}{2}$ in the large $U_c$ and $N$ limit.
As indicated by Eqs.~(\ref{eq:ne}) and (\ref{eq:nh}),
$a_{mk_>}$'s with $m = 1, ..., L$ correspond to electron excitations,
because they represent levels unoccupied initially but occupied in the intermediate state.
Similarly, $a_{mk_<}$'s with $m = L+1, ..., N$ correspond to hole excitations.
The expression of $a_{mk}=C_m/(\varepsilon_k-\varepsilon_m)$ indicates that
the squared amplitude of a state $k$ in a state $m$ decreases as their energy difference increases,
proportional to $(\varepsilon_k-\varepsilon_m)^{-2}$.
Therefore, if the intermediate one-electron energy $\varepsilon_m$ is well separated
from a band of $\varepsilon_k$, like the bound state $m = 1$, all $\varepsilon_k$ states
in the band would contribute almost equally to that intermediate state.
In contrast, if the intermediate state energy $\varepsilon_m$ is right next to
some $\varepsilon_k$'s, like the intermediate states in the continuum band,
those nearby $\varepsilon_k$ states contribute dominantly to that intermediate state.
Therefore,
as indicated by the long (blue) arrow between Figs.~\ref{fig:mi_discuss}(a) and~\ref{fig:mi_discuss}(b)
and the dotted line in Fig.~\ref{fig:mi_discuss}(a),
the intermediate bound state includes electron excitations
with an almost constant distribution function above the Fermi energy.
The intermediate state in the continuum includes electron or hole excitation
only if the intermediate state is right near the Fermi energy,
as indicated by the two short (pale blue and red) arrows
between Figs.~\ref{fig:mi_discuss}(a) and~\ref{fig:mi_discuss}(b),
giving rise to singular features.
This idea is further demonstrated by the (red) dashed line in Fig.~\ref{fig:mi_discuss}(b),
which represents $n_m^e=\sum_{k>} a_{mk>}^2$ for $m = 1,...,L$, the
levels with energies below zero, and
$n_m^h=\sum_{k<} a_{mk<}^2$ for $m=L+1,...,N$, the
levels with energies above zero.
The distribution functions $n^e_m$ and $n^h_m$ represent
how different {\it intermediate} eigenstates contribute to the
the e-h excitations, whereas
$n_{k_<}^h$ and $n_{k_>}^e$ represent
how different {\it initial} and {\it final} eigenstates contribute to the
the e-h excitations.

We apply a similar analysis to insulators,
as shown in Figs.~\ref{fig:mi_discuss}(c) and~\ref{fig:mi_discuss}(d).
The intermediate state DOS
is shown in gray (green) solid line in Fig.~\ref{fig:mi_discuss}(d).
It indicates two bound states, one below the bottom of the lower band
and the other within the gap.
Other $L-1$ and $N-L-1$ states are within the lower and upper continuum bands, respectively,
with energies shifted in a similar way as metals.
Therefore, the $L$ electrons fill right to the top of the lower band.
In the limit of $U_c, \Delta \gg W$, the two bound state energies are
$-\frac{U_c}{2}\pm \frac{1}{2}\sqrt{U_c^2+\Delta^2}$.
Further approximation in the limit of $U_c\gg \Delta$ gives rise to
the bound state energy below the bottom of the band,
$\varepsilon_{m=1}\cong -U_c-\Delta^2/(4U_c)$, and
the bound state energy inside the gap, $\varepsilon_{m=L+1} \cong \Delta^2/(4U_c)$.
The (red) dashed line in Fig.~\ref{fig:mi_discuss}(d) shows
$n_m^e$ for $m = 1, ..., L$ and $n_m^h$ for $m = L+1, ..., N$.
Because of the gap, the intermediate continuum states do not include
e-h excitation, and therefore,
the singularity in $n^e_{k_>}$ or $n^h_{k_<}$ disappears.
The states with substantial electron and hole excitations are
the bound states, $\varepsilon_{m=1}$ and $\varepsilon_{m=L+1}$, respectively,
as shown with the arrows between Figs.~\ref{fig:mi_discuss}(c) and~\ref{fig:mi_discuss}(d).
The shape of electron and hole distribution functions,
$n^e_{k_>}$ and $n^h_{k_<}$, are determined by the distance
between the intermediate bound states and the initially occupied or unoccupied bands.
The bound state below the lower band is well separated from the initially unoccupied band,
giving rise to an electron distribution function close to a constant.
In contrast, the bound state within the gap is close to the top of the initially occupied band,
which gives rise to a hole distribution function peaked near the gap.
This difference results in the strong asymmetry in the electron and hole distributions.

\begin{figure}
\begin{center}
\includegraphics[width=8.6cm]{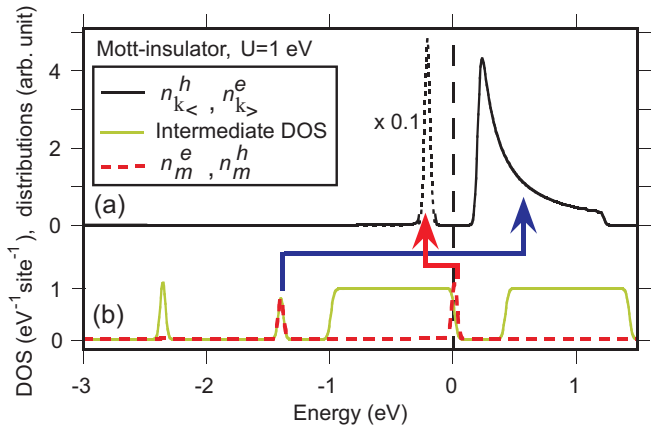}
\end{center}
\caption{\label{fig:mott_discuss} (color online)
(a) Occupation of holes and electrons for the Mott insulator with $U=1$ eV,
similar to the one shown in Fig.~\ref{fig:mott_rixs}(b).
For illustrative purpose, a smaller core-hole potential ($U_c=2$ eV)
is chosen.
(b) Intermediate DOS shown in gray (green) solid line and
$n^e_m$ and $n^h_m$, defined in the text, in (red) dashed line.
Arrows indicate how the two plots are related, as explained in the text.
}
\end{figure}

The results for the half-filled Mott insulators shown in Fig.~\ref{fig:mott_rixs}
are interpreted in a similar way.
The on-site Coulomb interaction splits the DOS into the lower and
upper Hubbard bands (LHB and UHB) with a gap proportional to $U$.
With two spin degrees of freedom on the site with the core hole,
two bound states are pulled below the bottom of the LHB,
as shown with the (green) solid line in Fig.~\ref{fig:mott_discuss}(b).
The lowest bound state at approximately $U_c$ below the LHB consists
mainly of the states from the LHB. The
second bound state, which is about $U$ higher in energy, is made of
mainly the states from the UHB.
For the lowest intermediate state, both bound states
are filled, giving an energy of $-2U_c+U$, which is equivalent to
the energy for two electrons at the site with a core hole.
After the de-excitation,
these bound electrons scatter to different states. However,
the LHB is already occupied in the ground state and the lowest bound state does not
contribute to the inelastic scattering, as indicated by $n_m^e$ and $n_m^h$
shown in the (red) dashed line in Fig.~\ref{fig:mott_discuss}(b).
In contrast, very few doubly
occupied states are present in the ground state and the second bound state leads to
the scattering to all the states in the UHB.
After filling the two
bound states, $N-2$ electrons are left to fill $N-1$ states
in the LHB in the intermediate state.
Therefore, the lowest intermediate state has a hole at the top of the LHB, and the
hole scatters only to a narrow region of the occupied DOS, as shown
in Figs.~\ref{fig:mott_discuss}(a) and \ref{fig:mott_discuss}(b).
The RIXS spectral line shape therefore
mainly reflects the unoccupied DOS, as is clear from the RIXS
spectra in Fig.~\ref{fig:mott_rixs}(a).

The relationship between the RIXS spectrum and the dynamic
structure factor is one of the interests in recent studies.
Our results on the forms of $n_{k_<}^h$ and $n_{k_>}^e$ for metals and insulators
allow us to find explicit expressions of $I(\omega)$ in
certain limits, which can be compared with the
dynamic structure factor $S(\omega)$ directly.
For example, in the limit of
 $U_c\gg \Delta,W$, we approximate
 $n^e_{k_>} \sim  U_c^{-2}$ for $\frac{\Delta}{2} \le \varepsilon_{k_>} \le \frac{\Delta + W}{2}$, and
 $n^h_{k_<} \sim (\varepsilon_{k_<})^{-2}$ for
 $-\frac{\Delta +W}{2} \le \varepsilon_{k_<} \le -\frac{\Delta}{2} $.
From these and Eq.~(\ref{eq:Inenh}), we find the RIXS
spectrum and its relation to the dynamic structure factor.
For $\Delta \le \omega <\Delta + \frac{W}{2}$, we obtain
\begin{equation}
I(\omega) \sim S(\omega)/\left[\frac{\Delta}{2}\left(\omega-\frac{\Delta}{2}\right)\right],
\end{equation}
where $S(\omega)\sim \omega-\Delta$.
For $\Delta+\frac{W}{2} \le \omega
<\Delta +W$,
\begin{equation}
I(\omega) \sim
S(\omega)/\left[\frac{W+\Delta}{2}\left(\omega-\frac{W+\Delta}{2}\right)\right],
\end{equation}
where $S(\omega)\sim \Delta+W-\omega$.
This shows that the relation between RIXS intensity and the dynamic structure factor
is not as straight-forward as expected from previous studies.
As for the integrated spectral weight,
we find that it  approaches to $AL(N-L)/N^2$
in the limit of $U_c \gg \Delta \gg W$,
recovering the results
found in Ref.~\onlinecite{Brink}.

Materials with narrow unoccupied and wide occupied bands provide
testing grounds for our results, since our theory predicts a narrow
RIXS spectrum with the hole excitations close to the Fermi level,
whereas a simple convolution of the unoccupied and occupied states
gives a wide RIXS spectrum.  Recent calculations~\cite{Inami}
suggest that the undoped manganite LaMnO$_3$ has such asymmetric
band widths (about 0.4 eV and 2 eV for the unoccupied and the
occupied bands, respectively) due to the layered $A$-type orbital ordering.
The RIXS spectrum measured for this material~\cite{Inami} indeed
shows a narrow peak with a small dispersion (less than 0.5 eV),
consistent with our theory. It would be also worthwhile to examine
the RIXS spectrum for materials with prominent features in the
occupied bands far below the Fermi level, which would distinguish
our theory from a simple convolution.
We further note that, when $\omega_{\rm in}$ is tuned to the
satellite (unscreened) peak
in the XAS spectrum, the situation is reversed:
a hole is bound to the core hole
and an electron is excited near the Fermi energy.
Such RIXS tuned at the satellite peak would probe occupied bands,
instead of empty bands.

\section{Summary}
In summary, we have shown that the spectral weight of charge
excitations in transition metal $K$-edge RIXS, where the core-hole
potential is larger than the effective bandwidth, is not determined
by the total energy of the shake-up, but results mainly from the
nature of the intermediate-state core-hole valence-shell-electron excitons.
We have identified the strong difference between excited electrons and holes,
which can be used to study occupied and unoccupied bands separately.
The methodology described in this work can be extended to more
complicated systems, for example, by incorporating results from {\it
ab initio} calculations, allowing a better interpretation of
transition metal $K$-edge RIXS.

\acknowledgments
 We thank John Hill, Stephane Grenier, and Jeroen
van den Brink for discussions. Work at Argonne National Laboratory
was supported by the U.S. Department of Energy, Office of Basic
Energy Sciences, under contract W-31-109-Eng-38. MvV was supported
by
 the U.S. Department of Energy (DE-FG02-03ER46097) and the Institute
for Nanoscience, Engineering, and Technology under a grant from
the U.S. Department of Education.


\begin{thebibliography}{0}

\bibitem{Kao} C.-C. Kao, W. A. L. Caliebe, J. B. Hastings, and
J.-M. Gillet, Phys. Rev. B {\bf 54}, 16 361 (1996).

\bibitem{Hill}
J. P. Hill, C.-C. Kao, W. A. L. Caliebe, M. Matsubara, A. Kotani, J. L. Peng, and R. L. Greene,
Phys. Rev. Lett. {\bf 80}, 4967 (1998).

\bibitem{Abbamonte} P. Abbamonte, C. A. Burns, E. D. Isaacs, P. M. Platzman, L. L. Miller, S. W. Cheong, and M. V. Klein,
Phys. Rev. Lett. {\bf 83}, 860 (1999).

\bibitem{Hasan}
M. Z. Hasan, E. D. Isaacs, Z.-X. Shen, L. L. Miller, K. Tsutsui, T. Tohyama, and S. Maekawa,
Science {\bf 288}, 1811 (2000);
M. Z. Hasan, P. A. Montano, E. D. Isaacs, Z.-X. Shen, H. Eisaki, S. K. Sinha, Z. Islam, N. Motoyama, and S. Uchida,
Phys. Rev. Lett. {\bf 88}, 177403 (2002).

\bibitem{Kim}
Y. J. Kim, J. P. Hill, C. A. Burns, S. Wakimoto, R. J. Birgeneau, D. Casa, T. Gog, and C. T. Venkataraman,
Phys. Rev. Lett. {\bf 89}, 177003 (2002);
Y.-J. Kim, J. P. Hill, H. Benthien, F. H. L. Essler, E. Jeckelmann,
H. S. Choi, T. W. Noh, N. Motoyama, K. M. Kojima, S. Uchida, D. Casa, and T. Gog,  {\it ibid.} {\bf 92}, 137402 (2004).

\bibitem{Doring} G. D\"oring,
C. Sternemann, A. Kaprolat, A. Mattila, K. H\"{a}m\"{a}l\"{a}inen, and W. Sch\"{u}lke, Phys. Rev. B {\bf 70},
085115 (2004).

\bibitem{Forte}
J. P. Hill, G. Blumberg, Y.-J. Kim, D. Ellis, S. Wakimoto, R. J.
Birgeneau, S. Komiya, Y. Ando, B. Liang, R. L. Greene, D.
Casa, and T. Gog, arXiv:0709.3274 (unpublished);
F. Forte, L. J. P. Ament, and J. van den Brink,
Phys. Rev. B {\bf 77}, 134428 (2008).

\bibitem{Saitoh}
E. Saitoh, S. Okamoto, K. T. Takahashi, K. Tobe, K. Yamamoto, T. Kimura, S. Ishihara, S. Maekawa, Y. Tokura,
Nature {\bf 410}, 180 (2001).

\bibitem{Kondo} H. Kondo, S. Ishihara, and S. Maekawa,  Phys.
Rev. B {\bf 64}, 014414 (2001).

\bibitem{Grenier}
S. Grenier, J. P. Hill, V. Kiryukhin, W. Ku, Y.-J. Kim, K. J. Thomas, S-W. Cheong, Y. Tokura, Y. Tomioka, D. Casa, and T. Gog,
Phys. Rev. Lett. {\bf 94}, 047203 (2005).

\bibitem{Gelmukhanov98} F. Gel'mukhanov and H. {\AA}gren
Phys. Rev. B {\bf 57}, 2780 (1998).

\bibitem{MvVgraphite}
M. van Veenendaal and P. Carra,  Phys. Rev. Lett. {\bf 78}, 2839 (1997).

\bibitem{Tsutsui} K. Tsutsui, T. Tohyama, and S. Maekawa,
 Phys. Rev. Lett. {\bf 83}, 3705 (1999); {\it ibid.} {\bf 91}, 117001 (2003).

\bibitem{Nomura} T. Nomura and J.-I. Igarashi,
J. Phys. Soc. Jpn. {\bf 73}, 1677 (2004);
T. Nomura and J.-I. Igarashi, Phys. Rev. B {\bf 71}, 035110 (2005).

\bibitem{Vernay}
F. Vernay, B. Moritz, I. S. Elfimov, J. Geck, D. Hawthorn, T. P. Devereaux,
and G. A. Sawatzky,
Phys. Rev. B {\bf 77}, 104519 (2008).

\bibitem{Brink} J. van den Brink and M. van Veenendaal,
J. Phys. Chem. Solids, {\bf 66}, 2145 (2005);
Europhys. Lett., {\bf 21} 121 (2006);
L. J. P. Ament, F. Forte, and J. van den Brink,
Phys. Rev. B {\bf 75}, 115118 (2007).

\bibitem{Ide}
T. Ide and A. Kotani, J. Phys. Soc. Jpn. {\bf  68}, 3100 (1999).

\bibitem{Laan81}
G. van der Laan, C. Westra, C. Haas, and G. A. Sawatzky,
Phys. Rev. B {\bf 23}, 4369 (1981);
P. J. W. Weijs, M. T. Czyzyk, J. F. van Acker, W. Speier,
J. B. Goedkoop, H. van Leuken, H. J. M. Hendrix, R. A. de Groot,
G. van der Laan, K. H. J. Buschow, G. Wiech, and J. C. Fuggle,
{\it ibid.} {\bf 41}, 11899(1990).

\bibitem{local.approx}
Even for a dispersionless core-hole band,
the momentum $\vec{k}$ is still a good quantum number because of the translational
symmetry of the system.
The core hole is directly coupled to the valence shell.
If we create a valence shell excitation with a momentum $\vec{q}$,
the core hole shifts also by a momentum $\vec{q}$.
In the end, the momentum transferred to the valence shell equals
the momentum lost by the photon.
Therefore, in principle, we can calculate the momentum dependence
by splitting the processes according to the momentum $\vec{q}$ of the excited e-h pairs.
The $K$-edge RIXS measures the integrated spectrum, which is calculated in this paper.

\bibitem{Davis79} L. C. Davis and L. A. Feldkamp, J. Appl. Phys.
{\bf 50}, 1944 (1979).

\bibitem{Davis80} L. A. Feldkamp and L. C. Davis, Phys. Rev. B
{\bf 22}, 4994 (1980).

\bibitem{Hubbard} J. Hubbard, Proc. Roy. Soc. London Ser. A {\bf 276} 238 (1963).

\bibitem{Mahan} G. D. Mahan, Phys. Rev. {\bf 163}, 612 (1967);
 P. Nozi\`eres and C. T. De Dominicis,
{\it ibid.} {\bf 178}, 1097 (1969).

\bibitem{Inami}
T. Inami, T. Fukuda, and J. Mizuki, S. Ishihara, H. Kondo,
H. Nakao, T. Matsumura, K. Hirota, Y. Murakami, S. Maekawa, and Y. Endoh,
Phys. Rev. B {\bf 67}, 045108 (2003).


\end{thebibliography}
\end{document}